\documentclass[12pt]{article}

\usepackage[dvips]{graphics}

\begin{document}
\title{Coherent dynamics of photoinduced nucleation processes}
\author{Kunio Ishida\\
Corporate Research and Development Center, Toshiba Corporation\\
1 Komukaitoshiba-cho, Saiwai-ku, Kawasaki 212-8582,
Japan\\
and\\
Keiichiro Nasu\\
Solid State Theory Division, Institute of Materials Structure Science, KEK,\\
Graduate University for Advanced Study, and CREST JST\\
1-1 Oho, Tsukuba, Ibaraki 305-0801, Japan}
\date{}

\maketitle

\begin{abstract}
We study the dynamics of initial nucleation processes of photoinduced
structural change of molecular crystals.
In order to describe the nonadiabatic transition in each molecule, we employ a model of localized electrons coupled 
with a fully quantized phonon mode, and the time-dependent
Schr\"odinger equation for the model is numerically solved.

We found a minimal model to describe the nucleation induced by injection of an
excited state of a single molecule in which multiple types of
 intermolecular interactions are required.
In this model coherently driven molecular distortion plays an
important role in the successive conversion of electronic states which
leads to photoinduced cooperative phenomena.
\end{abstract}

\section{Introduction}
\label{intro}

As the recent advance in laser technology has made it possible to
generate arbitrarily designed optical pulses, control of quantum mechanical states
of materials with those laser pulses has become to one of the central interests of
the research of future device applications.
For example, study on quantum information technology\cite{qi} has opened
up a new aspect of device applications, {\it i.e.}, utilization
of the phase(coherence) of the quantum mechanical states.
In those new fields, contrary to the conventional control methods of
electronic/vibrational states,  it is required to keep the coherence of the internal
states of materials during control processes, which means
that we need to understand the dynamics of quantum-mechanical
states in coherent regime.

On the other hand, it was also found in various materials that injection of photoexcited states 
induces cooperative phenomena regarding with the change of structural, magnetic, or
ferroelectric properties\cite{ct,pda,spin,binuclear,letard}.
These photoinduced cooperative phenomena are considered to have a common mechanism, 
and many experimental and/or theoretical studies have been presented
to make it clear\cite{nasu,ogawa,nasu2}.
In particular, when we are interested in controlling such
cooperativity by designed optical pulses,
it is necessary to understand the transient properties of the photoinduced
cooperative phenomena more deeply.
Thus, theoretical studies focused on the dynamics of the photoinduced 
cooperative phenomena are important at the current stage.

As we have shown in the previous papers\cite{ishida1,ishida2},
nonadiabaticity of electronic transitions is a key to understand
the temporal behavior of the above-mentioned phenomena.
Dynamics of nonadiabatic transitions has been studied since the
pioneering works by Landau\cite{landau} and
Zener\cite{zener}, and the bifurcation rate of wavefunction was analytically obtained in
general cases\cite{zhu}.
These studies mainly focused on the wavefunctions before/after
nonadiabatic transition,  and hence the time evolution of
wavefunctions itself is out of their scope.
On the other hand, the dynamics of nonadiabatic processes has been considered to be
important in, for example,  photochemical reactions\cite{chemical},
and hence computational methods of the dynamics have been proposed by many authors\cite{calc1}.
Since, however, those methods require the atomic coordinates/momenta to
be treated as 
classical variables due to the limited computational capacity at the
present time, they could discuss the wavefunctions after decoherence of
the atomic degrees of freedom takes place.
As a result they mentioned the reaction yield or the absorption rate after
various nonequilibrium processes.
On the contrary, the initial nucleation processes in photoinduced cooperative
phenomena involve consecutive switching of
potential energy surfaces (PESs) relevant to the dynamics of excited states,
and the coherence of the wavefunction of electrons/atoms should be taken
into account.
In other words, the wavefunction at every moment should be pursued to understand the
dynamics of the whole processes.
Hence, not only the bifurcation rate of the wavefunction
after nonadiabatic transition but also the wavefunction as a function of time
is required, which means that the 
conventional methods are not suitable for the theory of photoinduced cooperativity.

In this paper, we study the initial dynamics of the photoinduced
domain growth, which is characterized by nucleation processes before
decoherence of quantum-mechanical states takes place.
We also focus on the photoinduced phenomena in which electron-phonon
interactions play an important role and propose a minimal model to describe 
such cases.

The organization of the paper is as follows: in Section \ref{models}
the molecular model is introduced and the method of calculation is
described.
In Section \ref{results} the calculated results are shown. Section
\ref{disc}
is devoted to discussion and conclusions.

\section{models and method}
\label{models}

As we discussed in our previous papers\cite{ishida1,ishida2},
nonadiabatic transitions between quantized states are particularly
important to study the dynamical aspects of photoinduced cooperative phenomena.

In this paper, we focus on  the initial dynamics of a photoexcited state
in interacting molecules, fully quantizing the relevant vibration modes.
However, the dimension of the Hilbert space for the whole system
increases drastically by quantizing atomic variables, which means that numerical
calculation on those systems requires lots of computational
resources.
Thus, we employ a simplest model which is sufficient to describe the
photoinduced nucleation processes.
In the present model we consider molecules arrayed on a square lattice.
Electrons relevant to the nonadibatic transitions are assumed to be
localized in each molecule, and two electronic levels coupled
with a single vibration mode is taken into account per molecule.
The diabatic PESs with respect to the electronic
states in each molecule cross with each other and
that the nonadiabaticity in the dynamics is taken into account via ``spin-flip''
interaction between two electronic states.
This model is known as a simplest model to discuss the relaxation dynamics of, 
{\it e.g.}, photoisomerization of molecules\cite{mol}.
As for the intermolecular interaction, we take into account vibrational
coupling and the Coulomb interaction between excited state electrons.
$\beta$ affects to induce molecular distortion by the excited electrons in the adjacent
molecules which is also of the same order as the other interaction terms.

Hence, the Hamiltonian in the present study is described by:
\begin{eqnarray}
{\cal H} & =  & \sum_{\vec{r}} \left \{\frac{p_{\vec{r}}^2}{2}+
    \frac{\omega^2 u_{\vec{r}}^2}{2}+ ( \sqrt{2\hbar
      \omega^3}sq_{\vec{r}}+ \varepsilon \hbar \omega + s^2 \hbar
    \omega ) \hat{n}_{\vec{r}}+\lambda \sigma_x^{\vec{r}} \right
\}\nonumber \\
& + & \sum_{\langle \vec{r},\vec{r'}\rangle} [ \alpha \omega^2
     (u_{\vec{r}}-\beta\hat{n}_{\vec{r}})(u_{\vec{r'}}-\beta\hat{n}_{\vec{r'}}) - \{ V - W (u_{\vec{r}}+u_{\vec{r'}}) \} \hat{n}_{\vec{r}} \hat{n}_{\vec{r'}} ],
\label{ham}
\end{eqnarray}
where $p_{\vec{r}}$ and $u_{\vec{r}}$ are the momentum and coordinate
operators for the vibration mode of a molecule at site $\vec{r}$, respectively.
The second sum which gives the intermolecular interaction is taken over
all the pairs on nearest neighbor sites, where the Coulomb interaction
between excited state electrons are modified by molecular distortion.
The vibrational period of an individual molecule is denoted by $T=2\pi/\omega$ in
the rest of the paper.

\begin{figure}[h]
\begin{center}
\scalebox{0.7}{\includegraphics*{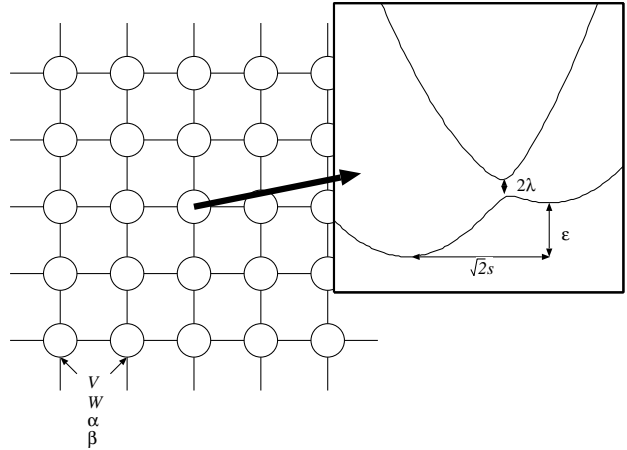}}
\caption{Schematic view of the model. Circles denote the molecules
  with two electronic states and a vibrational mode. Adiabatic potential energy
  surfaces for an individual molecule is shown in the inset.}
\label{schematic}
\end{center}
\end{figure}

A schematic view of the present model is shown in Fig.\
\ref{schematic}.
The two diabatic PESs for an
individual molecule are crossed with each other, and the nonadiabatic
coupling constant $\lambda$ acts to separate them into two adiabatic
PESs.
We chose the values of the parameters as:
$\varepsilon=1.6$,$s=1.4$,$V=1.1$,$W=0.2$, $\alpha=0.1$, $\beta=0.2$, and $\lambda=0.2$.
Although those values are typical for organic molecules as for
electron-vibration coupling\cite{wp} and  the intermolecular Coulomb interaction\cite{mol},
the other parameters are not easy to determine their values either from
theoretical calculations or experimental results.
We only mention that the order of magnitude for the parameters is
estimated referring to those for typical organic materials.
The electronic states at site $\vec{r}$ are denoted by $|\downarrow \rangle_{\vec{r}}$ (ground
state) and $|\uparrow \rangle_{\vec{r}}$ (excited state) and
$\sigma_i^{\vec{r}}\ (i=x,y,z)$ are the Pauli matrices which act only on
the electronic states of the molecule at site $\vec{r}$.
$\hat{n}_{\vec{r}}$ denotes the density of the electron in $|\uparrow
\rangle_{\vec{r}}$ which is rewritten as $\hat{n}_{\vec{r}}=\sigma_z^{\vec{r}}+1/2$.
The model and the notations of the parameters are schematically shown in
Fig.\ \ref{schematic}.

Details of the quantization procedure of the vibration mode of each molecule will be published elsewhere.
We mention that the basis set for the vibronic states is composed of the Fock states
shown in Ref.\ \cite{old}.
The phonon dispersion relation of the vibration mode is given by 
\begin{equation}
\Omega(\vec{k})= \omega \sqrt{1+2\alpha (\cos k_x+\cos k_y)},
\label{disp}
\end{equation}
where $(k_x, k_y)$ denote the reciprocal lattice vector of
the square lattice, and the lattice constant is taken to be unity.
The quantized states on each diabatic PES of a single molecule
are the vibronic states $| n \sigma \rangle_{\vec{r}}$ ($n =
0,1,2,...$, $\sigma=\uparrow,\downarrow$) in the Fock
representation, where the coordinate of the molecule is labelled by
$\vec{r}$.
$|n \uparrow \rangle$ is related with $|n
\downarrow \rangle$ by
\begin{equation}
| n \uparrow \rangle = |\uparrow \rangle \langle \downarrow
|e^{s(a^\dagger + a)} | n \downarrow \rangle,
\end{equation}
where $e^{s(a^\dagger +a)}$ denotes the translation operator in the
vibration coordinate space\cite{wp2}.
We note that this Ising-like model is similar to the one to study the thermodynamical
properties of the Jahn-Teller effect\cite{jahn}, though the nonequilibrium dynamics of the excited states in the model
has not been understood.

We obtain the numerical solution of the time-dependent Schr\"odinger equation for the
Hamiltonian (\ref{ham}) by the Runge-Kutta method.
In each series of calculations, one of the molecules on 128$\times$128 lattice
is initially in the
Franck-Condon state, while the others are in the ground state, which
corresponds to the injection of a photoexcited state to a single
molecule at the origin.
In solving the Schr\"odinger equation, we applied a mean-field
approximation in which the contribution of the wavefunction at the
nearest neighbor sites is substituted by the average value with
respect to the wavefunction $|\Phi(t)\rangle$.
The detail of the approximation is described in the appendix.
We only mention here that this approximation is equivalent to decomposing the wavefunction of the total
system $|\Phi(t) \rangle$ into a product of the wavefunctions at each
molecule,
{\it i.e.}, 
\begin{equation}
|\Phi(t) \rangle = |\phi(t) \rangle_{\vec{r_1}} \otimes |\phi(t)
\rangle_{\vec{r_2}} \otimes ... |\phi(t) \rangle_{\vec{r_N}},
\label{decompose}
\end{equation}
where $N$ denotes the number of molecules in the system.
Thus, we can solve the differential equation for each molecule
when only the average values of the properties for adjacent molecules are provided,
which means that the present calculation method is suitable for parallel
computing.
Hence, we have made it possible to handle more than 10000 molecules by
the present method.

\section{calculated results}
\label{results}

\begin{figure}
\begin{center}
\scalebox{0.35}{\includegraphics*{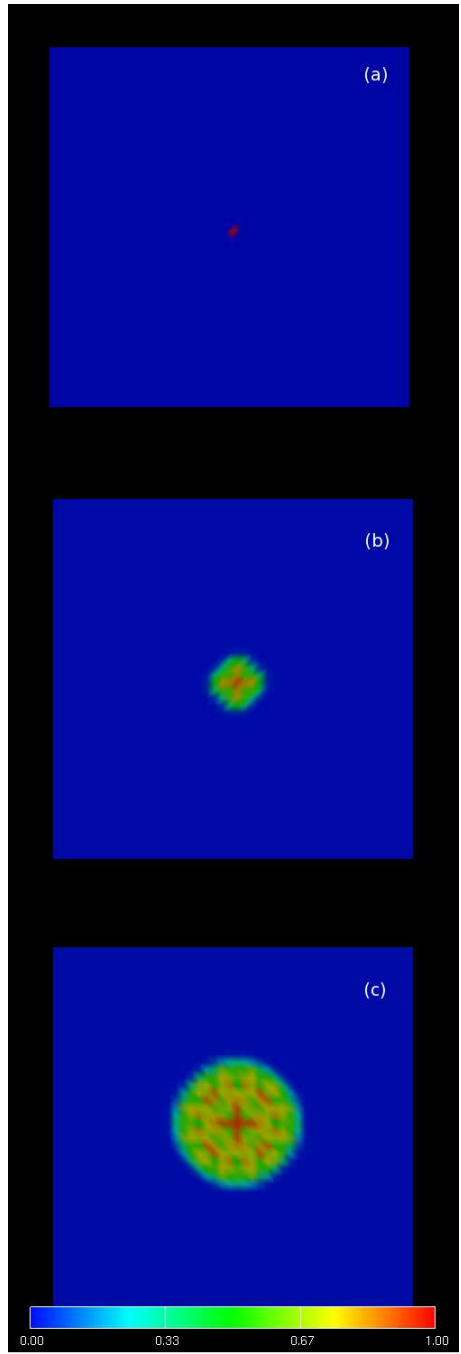}}
\caption{Population of the excited electronic state $N(\vec{r},t)$ on
  $48 \times 48$ lattice for
  (a) $t=0$, (b) $t=5T$, and (c) $t=10T$.}
\label{pop1}
\end{center}
\end{figure}

We have pointed out that the population of the excited electronic state $|\uparrow
\rangle_{\vec{r}}$ is suitable for understanding the dynamics of the initial nucleation processes
\cite{ishida1,ishida2}.
Thus, we show in Fig.\ \ref{pop1} the population of $|\uparrow
\rangle_{\vec{r}}$ for 48$\times$48 sites around the injected excited state (``seed'') defined by
\begin{equation}
N(\vec{r},t) = \langle \Phi(t) | \hat{n}_{\vec{r}} | \Phi(t)
\rangle,
\end{equation}
for $t=0$, $5T$, and $10T$.
Figure \ref{pop1} shows that the number of molecules in the excited
electronic state increase surrounding the initially excited molecule.
Those molecules will constitute a photoinduced domain observed in many experiments\cite{ct,pda,spin,binuclear,letard}
and thus the present calculation described the initial processes of
the photoinduced cooperative phenomena, {\it i.e.} photoinduced
nucleation triggered by an injected excited state.

\begin{figure}
\begin{center}
\scalebox{0.35}{\includegraphics*{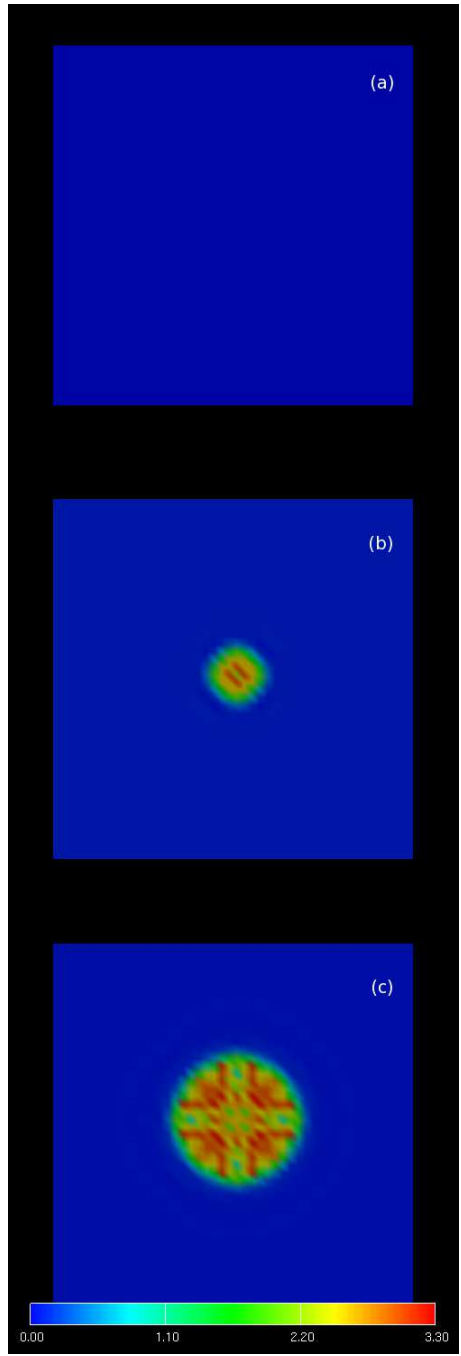}}
\caption{Molecular distortion $\zeta(\vec{r},t)$ on $48 \times 48$
  lattice for
  (a) $t=0$, (b) $t=5T$, and (c) $t=10T$.}
\label{lat1}
\end{center}
\end{figure}

Since the population transfer is induced by adiabatic transition on
the PESs in Fig.\ \ref{schematic}, molecular distortion
is also relevant to the nucleation processes.
Hence, molecular distortion defined by
\begin{equation}
\zeta (\vec{r},t) = \langle \Phi(t) |u_{\vec{r}} |\Phi(t) \rangle,
\end{equation}
is calculated to discuss different aspects of the nucleation processes.
Figure \ref{lat1} shows $\zeta (\vec{r},t)$ for $t=0$, $5T$, and $10T$.
We found that the molecular distortion is coherently driven in the system
and that the vibrational energy is propagated by coherent phonons.

\begin{figure}
\begin{center}
\scalebox{0.7}{\includegraphics*{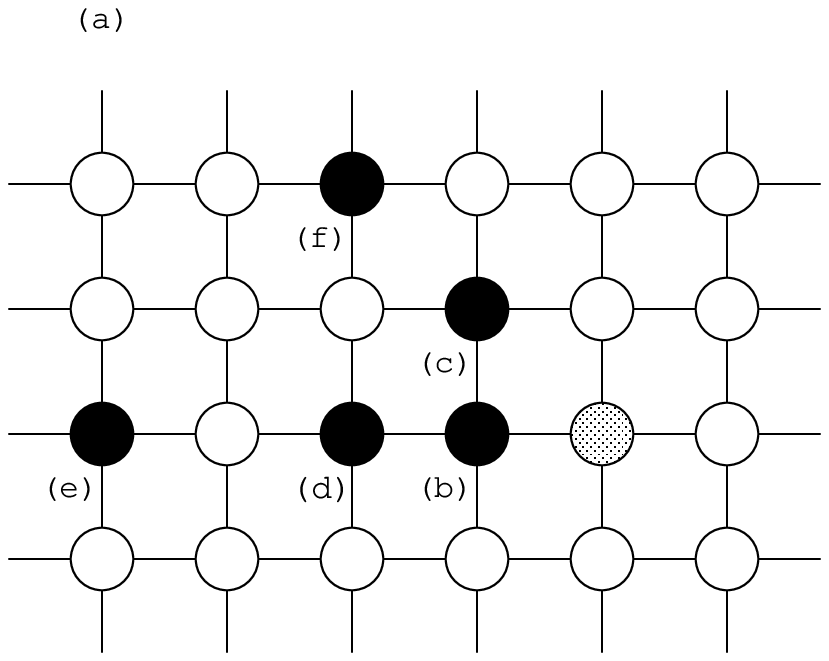}}
\scalebox{0.5}{\includegraphics*{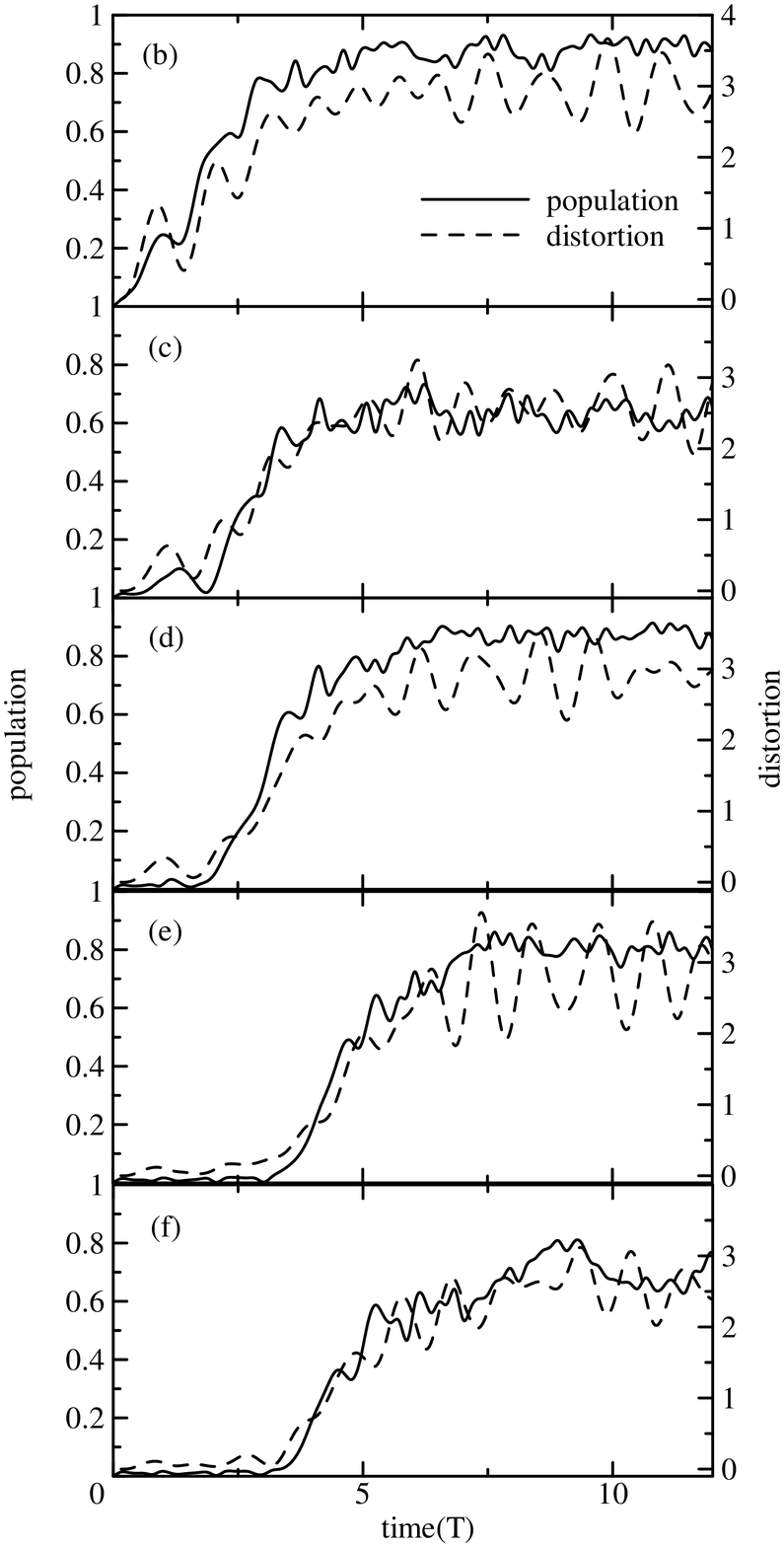}}
\caption{Time-dependence of $N(\vec{r},t)$ and $\zeta
  (\vec{r},t)$. $\vec{r}$ for each figure corresponds to the filled circles
  in the top figure, and the ``seed'' is denoted by the shaded circle.}
\label{td}
\end{center}
\end{figure}
The elementary processes of the nucleation is 
understood by comparing $N(\vec{r},t)$ and $\zeta (\vec{r},t)$ as
functions of time.
Although these properties seem to have similar temporal behavior at 
first glance, we found
that the difference in their initial growth dynamics is a key to
understand the details of the nucleation processes.
We show in Figs.\ \ref{td}-(b)-(f) $N(\vec{r},t)$ and $\zeta (\vec{r},t)$ as
functions of time, where $\vec{r}$ corresponds to the molecular sites shown 
by filled circles in Fig.\ \ref{td}-(a).
As we mentioned, the boundary between the converted (excited state) domain and the ground state
domain moves to extend the former in the system.
Thus the values of $N(\vec{r},t)$ and $\zeta(\vec{r},t)$ rise at later time as
the distance to the ``seed'' increases.
Comparing Figs.\ \ref{td}-(b)-(f), we found that the distance to
the ``seed'' should refer to the Manhattan distance in discussing
the time of rising of those values when $|\vec{r}|$ is small.
However, as the growth process proceeds, the number of the excited
molecules increases and  the conversion of the electronic states takes place as
 in the continuous systems and the boundary between domains becomes a
 circle (Fig.\ \ref{lat1}-(c)).
Hence, the Euclidean distance becomes appropriate for the distance between two molecules.
This behavior is reminiscent of the propagation of coherent phonons,
and the nucleation processes are driven by those coherent motion
of the molecules.

Figures \ref{td}-(b)-(d) also show that an oscillating component in both $N(\vec{r},t)$ and $\zeta(\vec{r},t)$ 
appears before transition to $|\uparrow \rangle_{\vec{r}}$ is realized.
We found that, after $N(\vec{r},t)$ and/or $\zeta(\vec{r},t)$ begins
to grow, it takes longer time to complete electronic state conversion in the molecules close to the ``seed'' than in the others.
To be more precise, a precursor to the electronic state conversion is 
observed in molecules at the nearest neighbor and the next nearest neighbor of the ``seed''.
This behavior is understood as a preliminary process required to overcome potential energy barrier 
at the beginning of nucleation.
Thus, in the present model, the first process of the nucleation
corresponds to making a cluster of the $\sim 10$ converted molecules around the ``seed.''
Then the other molecules will suffer electronic state conversion which
 smoothly takes place as Figs.\ \ref{td}-(e) and (f) show.

The above properties are relevant to the mechanism of the photoinduced
cooperative phenomena, {\it i.e.}, each molecule in the ground state receives
energy through the vibrational coupling $\alpha$ first and the
molecule begins to vibrate.
Then, the nonadiabatic coupling $\lambda$ induces population transfer
with the assistance of the Coulomb interaction $V$, and thus the molecule
turns to belong to the converted domain. 
Once each molecule belongs to the converted domain, motion of $N(\vec{r},t)$ and $\zeta
(\vec{r},t)$ almost disappear and it does not return to the ground
state during the current simulation, since intermolecular interactions
make it remain in the excited state domain.
We also point out that the intermolecular Coulomb interaction $V$ enhances the population
transfer rate particularly when the number of adjacent molecules  in the
excited state increases.

Figure \ref{td}-(b) also shows that the molecular distortion does
not grow for $\sim 3T$ after photoexcitation.
Hence, it takes $\sim 4T$ for the population transfer to be completed even
for the molecules in the nearest neighbor sites.
Since the population transfer reflects on the electronic/optical properties 
of the molecules, the change of those physical properties takes place
$\sim 400-800$ fsec ($T$ is $\sim 100-200$fsec in typical organic
molecular systems) after photoexcitation.

\begin{figure}
\begin{center}
\scalebox{0.5}{\includegraphics*{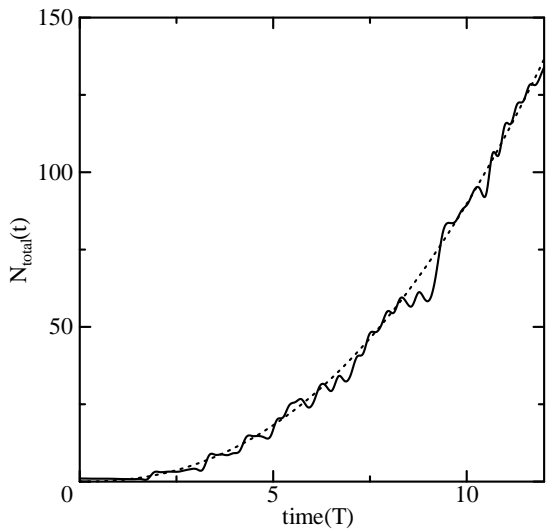}}
\caption{$N_{\rm total}(t)$ as a function of time. The dotted line 
  proportional to $t^{2.3}$ is drawn as a guide for the eyes.}
\label{number}
\end{center}
\end{figure}
The sum of the excited state population $N_{\rm total}(t)=\sum_{\vec{r}} N(\vec{r},t)$
indicates the measure for the growth rate of the photoinduced domain.
Figure \ref{number} shows that $N_{\rm total}(t)$ increases as $\sim t^{2.3}$.
except in the very first stage of the domain growth.
As a result we obtain that the radius of the photoinduced domain
behaves as $\sim t^{1.2}$, which is understood by the picture that  
the growth of the domain is predominantly driven by propagation of coherent phonons 
rather than diffusion processes.
As the vibrational coherence is lost, diffusion process becomes more
important and the domain growth will slow down to make the radius of the
domain increase as $\sim \sqrt{t}$.
Since vibrational coherence survives for a few psec in typical organic
molecules\cite{wp}, the present calculation is valid only in the 
time range studied in this paper, and the decoherence of the vibrational
states should be taken into account to study the growth dynamics of the
photoinduced domain in a longer time scale.

\section{discussion and conclusions}
\label{disc}

In this paper we study the coherent dynamics of photoinduced
nucleation processes in organic molecular systems.
When a single molecule excited to the Franck-Condon state, it induces
distortion of adjacent molecules, and the excitation energy is
transferred to the other molecules coherently.
Once the molecules start to vibrate, the electronic state conversion
from $|\downarrow \rangle_{\vec{r}}$ to $|\uparrow \rangle_{\vec{r}}$ takes place and thus 
photoinduced domain grows.
This is the basic scenario of the initial photoinduced nucleation
processes where coherent phonons play an important role.
In fact, the size of the converted domain (diameter) is almost linearly increases as the
nucleation proceeds, which shows that energy diffusion is
subsidiary in the initial processes.
However, as the decoherence of vibrational states takes place,
excitation energy propagation in the system will be dominated by 
diffusion processes, and hence the growth rate will be $\propto \sqrt{t}$
after all.
We mention that these properties will be reflected on the time-resolved
spectra of {\it e.g.}, reflectance, absorbance, or Raman scattering
intensity and that the ultrafast spectroscopy
will give a key to understand the coherent nature of the nucleation processes.

We stress that, contrary to the one-dimensional case\cite{ogawa}, the domain growth by successive conversion of the molecules
 does not take place in higher dimensional cases if
intermolecular interactions other than vibrational coupling $\alpha$ is neglected.
This is a particular property in higher dimensional systems, since it
was pointed out that the converted domain grows only by the
intermolecular interaction between molecular distortion\cite{ogawa}.
In general, the value of $\alpha$ should be less $1/M$ where $M$ is the coordination number of the lattice.
Hence, the maximum value of $\alpha$ is smaller as the dimensionality
of the system is higher.
The present study shows that, even in two-dimensional systems, $\alpha$
is not sufficient to induce cooperative phenomena, and thus other
interactions such as Coulomb interaction between electrons are necessary.
Since $M$ is larger in higher dimensional systems, this result is applicable
to three-dimensional systems, and thus we conclude that the
photoinduced nucleation processes are realized by cooperation of
various types of intermolecular interactions in general cases.
We mention that the results described in this paper are independent of the model and
the values of the parameters for which the nucleation takes place.

In the present paper, we assume that only a single relevant
vibration mode exists in each molecule.
However, the nonadiabatic transition within a single molecule is
strongly affected by the structure of the PESs.
In particular, when multiple vibration modes are taken into account, 
the dynamics of the nucleation processes depends on the topological structure of the intersections of the PESs, {\it e.g.},
existence of conical intersections. 
Hence, {\it ab initio} electronic-structure calculations of specific materials are 
important for a detailed discussion of such material-dependent features of the nucleation
processes, and the dynamics calculation in the present paper
should be combined with those electronic-structure calculations in the future.
We, however, stress that the present results give the basic properties
of the nucleation dynamics in coherent regime and that the qualitative
feature of the domain growth is sufficiently discussed in this paper.

In order to control the nucleation dynamics by outer field, {\it
  e.g.}, laser pulses, we should estimate the effect of decoherence of the
quantum-mechanical states.
We point out that it is possible to take into account the decoherence by
embedding the system in a large 'reservoir' and by tracing out the
dynamical variables regarding with the reservoir.
We also stress that such studies will contribute to realizing the coherent control of the
photoinduced domains, which will be important both from a physical
point of view and  device applications\cite{cc}.

\noindent {\bf acknowledgments}

One of the authors(K.I.) thanks K. Takaoka and H. Asai for helpful advice. 
This work was supported by the Next Generation Super Computing Project,  
Nanoscience Program, MEXT, Japan, and the numerical calculations were carried out on the computers at the
Research Center for Computational Science, National Institutes of Natural
Sciences.

\appendix

\section{equations of motion for numerical calculations}
The Schr\"odinger equation for the system is written by
\begin{equation}
i\frac{\partial}{\partial t} | \Phi(t) \rangle = {\cal H} |\Phi (t)
\rangle.
\label{eom1}
\end{equation}
When the wavefunction $|\Phi(t) \rangle$ is decomposed by those for
individual molecules as shown in Eq.\ (\ref{decompose}), each
component $|\phi(t) \rangle_{\vec{r}}$ obeys the following equation:
\begin{eqnarray}
i\frac{\partial}{\partial t}|\phi (t) \rangle_{\vec{r}} & = & 
 \left \{\frac{p_{\vec{r}}^2}{2}+ \frac{\omega^2 u_{\vec{r}}^2}{2}+ (
     \sqrt{2\hbar \omega^3}sq_{\vec{r}}+ \varepsilon \hbar \omega +
     s^2 \hbar \omega ) \hat{n}_{\vec{r}}+\lambda \sigma_x^{\vec{r}}
 \right \} |\phi(t) \rangle_{\vec{r}}   \nonumber \\
&  + & \sum_{\vec{r'}} [ \alpha \omega^2
     (u_{\vec{r}}-\beta\hat{n}_{\vec{r}})(u_{\vec{r'}}-\beta\hat{n}_{\vec{r'}}) - \{ V - W (u_{\vec{r}}+u_{\vec{r'}}) \} \hat{n}_{\vec{r}} \hat{n}_{\vec{r'}} ]|\phi(t) \rangle_{\vec{r}},
\label{eom2}
\end{eqnarray}
where the sum is taken over the nearest neighbor sites of $\vec{r}$.

We apply a mean-field approximation to Eq.\ (\ref{eom2}), {\it i.e.},
the effects of molecules in the nearest neighbor sites are substituted
by their average values.
Hence, after canonical quantization of the $p_{\vec{r}}$ and $u_{\vec{r}}$,
the following equations of motion for each molecule are derived:
\begin{eqnarray}
i \frac{\partial}{\partial t}|\phi (t) \rangle_{\vec{r}} & = & 
\hbar \omega [ a_{\vec{r}}^\dagger a_{\vec{r}} + 
    \{ s (a_{\vec{r}}^\dagger + a_{\vec{r}} )+ \varepsilon + s^2 \}
    \hat{n}_{\vec{r}}+\lambda \sigma_x^{\vec{r}} ] |\phi (t) \rangle_{\vec{r}}
    \nonumber \\
&  + & \sum_{\vec{r'}} \left [ \alpha \omega^2
      \{
         \gamma (a_{\vec{r}}^\dagger+a_{\vec{r}})-\beta
         \hat{n}_{\vec{r}} \}(\langle u_{\vec{r'}}\rangle 
-  \beta\langle \hat{n}_{\vec{r'}}\rangle )  \right .  \nonumber \\
& - & \left .  \{ V - W
      \{
         \gamma (a_{\vec{r}}^\dagger+a_{\vec{r}})
          +\langle u_{\vec{r'}}\rangle  \} \} \hat{n}_{\vec{r}}
     \langle \hat{n}_{\vec{r'}}\rangle \right ]|\phi(t)
 \rangle_{\vec{r}},
\label{eom3}
\end{eqnarray}
where
$a_{\vec{r}}^{(\dagger)}$ denotes the annihilation (creation) operator
of the vibration mode and $\gamma = \sqrt{\hbar/2\omega}$.
Intermolecular interaction is taken into account through the average
value over the decomposed wavefunction and
\begin{equation}
\langle X_{\vec{r}} \rangle_{\vec{r}} =  \ _{\vec{r}}\! \langle \phi(t) | X_{\vec{r}} | \phi(t)
\rangle_{\vec{r}}.
\end{equation}

\end{document}